\documentclass[12pt,a4paper]{article}
\usepackage[T1]{fontenc}        
\usepackage{lmodern}            
\usepackage[british]{babel}
\usepackage{amsmath}
\usepackage{amssymb}
\usepackage{amsfonts}
\usepackage{dsfont}
\usepackage{setspace}           
\usepackage[dvips]{graphicx}
\title{
  {\vspace{-2cm} \normalsize
     \includegraphics[width=80mm]{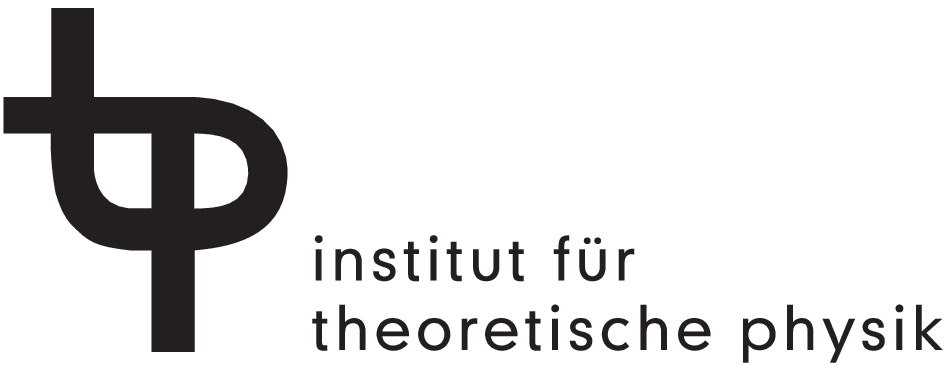}
     \hfill\parbox[b][30mm][t]{35mm}{MS-TP-11-05}%
  }\\[25mm]
Twisted mass chiral perturbation theory for $2+1+1$ quark flavours}
\author{Gernot M\"unster and Tobias Sudmann%
        \thanks{present address: Deloitte \& Touche GmbH, 60486 Frankfurt}\\
        Institut f\"ur Theoretische Physik,
        Universit\"at M\"unster\\
        Wilhelm-Klemm-Str.~9, D-48149 M\"unster, Germany\\
        e-mail: munsteg@uni-muenster.de}
\date{March 8, 2011}
\newcommand{\I}{\ensuremath{\mathrm{i}}}
\newcommand{\E}{\ensuremath{\mathrm{e}}}
\newcommand{\Tr}[1]{\ensuremath{\operatorname{Tr}\left(#1\right)}}

\begin{document}
\maketitle

\begin{abstract}
We present results for the masses of pseudoscalar mesons in twisted mass
lattice QCD with a degenerate doublet of $u$ and $d$ quarks and a
non-degenerate doublet of $s$ and $c$ quarks in the framework of
next-to-leading order chiral perturbation theory, including lattice effects
up to $\mathcal{O}(a^2)$. The masses depend on the two twist angles for the
light and heavy sectors. For maximal twist in both sectors,
$\mathcal{O}(a)$-improvement is explicitly exhibited.  The mixing of
flavour-neutral mesons is also discussed, and results in the literature for
the case of degenerate $s$ and $c$ quarks are corrected.
\\[5mm]
\end{abstract}
\section{Introduction}

For non-perturbative studies of Quantum Chromodynamics on a space-time
lattice the formulation with a chirally rotated mass term \cite{TM1,TM2} has
proven to be a very effective framework. This so-called twisted mass lattice
QCD implies automatic $\mathcal{O}(a)$ Symanzik-improvement if the twist
angle is set to a value of $\pi/2$ \cite{FR1,FR2}. Numerical simulations of
twisted mass lattice QCD have allowed to determine a number of physically
relevant hadronic parameters, notably the Gasser-Leutwyler low-energy
constants, to a high precision.

First studies of twisted mass lattice QCD have been made with two
mass-degenerate quark flavours, representing the up and down quarks, and
have allowed to obtain results for quantities in the pion sector of QCD
\cite{etmc2-1,etmc2-2,etmc2-3}, for a review see \cite{Shindler}. In recent
work these calculations have been extended to $2+1+1$ quark flavours,
meaning a degenerate doublet of up and down quarks and a non-degenerate
doublet of charm and strange quarks \cite{FR3,etmc1,etmc2,etmc3}.  The
chiral twist is implemented with two independent twist angles in the up-down
sector and in the strange-charm sector. Both of these angles are tuned to
$\pi/2$. In the numerical simulations the mass of the charm quark is on the
cut-off scale, whereas the mass of the strange quark is near its physical
value, so that physical results about observables in the sector of the the
three lightest quarks can be obtained.

For the analysis of the numerical data, which are obtained at varying quark
masses, chiral perturbation theory is an invaluable tool, for a review see
\cite{Sharpe}. Chiral perturbation yields analytical formulae for the
dependence of physical quantities like meson masses and decay constants on
the quark masses. It amounts to an expansion around the chirally symmetric
limit of vanishing quark masses, and is reliable for sufficiently small
quark masses.

For the case of twisted mass lattice QCD, chiral perturbation theory has
first been developed and used for $N_{\textrm{f}} = 2$ quark flavours
\cite{MS,Muenster,Sco,Sharpe-Wu2}. It has been extended to $N_{\textrm{f}} =
3$, including the strange quark, in \cite{Mue-Sud-1}.  In view of the
numerical work on $N_{\textrm{f}} = 2+1+1$ twisted mass lattice QCD, it is
the purpose of this article to present chiral perturbation theory for this
situation.

Of course, at this point the question of the region of applicability of
chiral perturbation arises. Chiral perturbation theory is well applicable to
quark masses as light as those of the up and down quarks. The mass of the
strange quark has turned out to at the border of the region of validity of
chiral perturbation theory, depending on the order of the expansion and the
use of partial resummations. The charm quark is definitely too heavy to be
treated within chiral perturbation theory.

Therefore, chiral perturbation for twisted mass lattice QCD with
$N_{\textrm{f}} = 2+1+1$ quark flavours is not applicable to quantities in
all 4 quark sectors. In this article we restrict ourselves to quantities in
the pion- and the kaon-sector. Moreover, for physical values of the strange
quark mass, the formulae cannot be applied to obtain precise quantitative
results in the kaon sector. Instead, the meaning of these calculations is
different. Primarily, the idea is to explicitly reveal the structure of
$N_{\textrm{f}} = 2+1+1$ twisted mass lattice QCD in the sense of its
dependence on the two independent twist angles. Chiral perturbation displays
how the values of the two twist angles affect the observables in the
different mesonic sectors, and explicitly exhibits the requirements for
automatic $\mathcal{O}(a)$ improvement. Secondly, chiral perturbation theory
yields the formulae that would allow to extract $N_{\textrm{f}} = 3$
low-energy constants from data in the region of unphysically small strange
quark masses.

In this letter we present the results of chiral perturbation theory for
$N_{\textrm{f}} = 2+1+1$ twisted mass lattice QCD in next-to-leading order.
Lattice effects are included up to order $a^2$. The special case
$N_{\textrm{f}} = 2+2$ with two degenerate quark doublets has already been
considered in \cite{ALWW}. In next-to-leading order our results show
differences to \cite{ALWW}, see below.

\section{Structure of the theory}

\subsection{Field definitions}

Chiral perturbation theory is formulated in terms of the pseudo-Goldstone
fields for spontaneously broken $\mathrm{SU}(N_{\textrm{f}})_L \otimes
\mathrm{SU}(N_{\textrm{f}})_R$ chiral symmetry. For $N_{\textrm{f}} = 4$,
the 15 canonical fields $\phi_a(x)$, $a=1 \ldots 15$, corresponding to the
generators $\lambda_a$ of SU(4) (see App.~\ref{generators}), are assembled
in the matrix valued field $\lambda_a \phi_a(x)$. Referring to the usual
particle multiplets, the canonical fields are also denoted
$(\phi_a)=(\vec{\pi}, \vec{K}, \eta_8, \vec{D}, \vec{D}_s, \eta_{15})$. The
fields corresponding to the particle eigenstates are given by the linear
combinations
\begin{align}
\pi^\pm       &= \tfrac{1}{\sqrt2} (\pi_1\mp \I\pi_2)\,, &
\pi^0         &= \pi_3\,,\nonumber\\
K^\pm         &= \tfrac{1}{\sqrt2} (K_4\mp \I K_5)\,, &
K^0,\bar{K}^0 &= \tfrac{1}{\sqrt2} (K_6\mp \I K_7)\,,\\
D^\pm         &= \tfrac{1}{\sqrt2} (D_{11}\pm \I D_{12})\,, &
D^0,\bar{D}^0 &= \tfrac{1}{\sqrt2} (D_9\pm \I D_{10})\,, &
D_s^\pm       &= \tfrac{1}{\sqrt2} (D_{s,13}\pm \I D_{s,14})\,.\nonumber
\end{align}
The field configuration matrix then reads
\begin{equation}\label{eq:fields}
\frac1{\sqrt2}\lambda_a \phi_a =
\begin{pmatrix}
\frac1{\sqrt2}\pi^0 & \pi^+                  & K^+   & \bar D^0\\
\pi^-               & \frac{-1}{\sqrt2}\pi^0 & K^0   & D^-\\
K^-                 & \bar K^0               & 0     & D_s^-\\
D^0                 & D^+                    & D_s^+ & 0
\end{pmatrix}
+ \frac1{\sqrt2} [\lambda_8\eta_8 + \lambda_{15}\eta_{15}]\,.
\end{equation}
It enters chiral perturbation theory in terms of the exponential
parameterisation $U(x)=\exp\big[\tfrac{\I}{F_0}\,\lambda_a\phi_a(x)\big]$,
where $F_0$ is a low-energy constant. The effective Lagrangian for $U(x)$
contains chiral symmetry breaking quark mass terms and lattice terms. To
leading order (LO) it is given by
\begin{equation}
\mathcal{L}_\mathrm{LO}
= \frac{F_0^2}{4} \Tr{\partial_{\mu} U \, \partial_{\mu} U^\dagger }
- \frac{F_0^2}{4} \Tr{\chi U^\dagger + U \chi^\dagger}
- \frac{F_0^2}{4} \Tr{\rho \, U^\dagger + U \rho^\dagger}.
\end{equation}
Here the quark mass term is
\begin{equation}
\chi = 2B_0 M
\end{equation}
with the quark mass matrix $M$,
and the lattice artifacts are represented by
\begin{equation}
\rho = 2 W_0 a \mathds{1}\,,
\end{equation}
where $a$ is the lattice spacing and $B_0$ and $W_0$ are additional
low-energy parameters. In next-to-leading order (NLO) the contributions to
the effective Lagrangian contain the Gasser-Leutwyler coefficients $L_i$, as
well as new coefficients $W_i, W'_i$, parameterising the lattice artifacts.
We refer the reader to \cite{Sharpe-Wu2,Mue-Sud-1} for definitions and
details about the NLO Lagrangian.

\subsection{Mass terms}

In the lattice regularisation, the untwisted physical quark mass $m'$ is
proportional to the difference $m_0-m_{0c}$ of the bare and the critical
mass; the prime indicates that the quark mass shift already includes an
${\cal O}(a)$ correction $aW_0/B_0$, see \cite{Sharpe-Wu2}. In the
$(u,d)$-sector the twisted mass $\mu$ is conveniently introduced via the
term $\I\mu\tau_3$. In chiral perturbation theory we use the variables
\begin{equation}
\chi' = 2B_0(m'+\I\mu\tau_3) = 2B_0 m_q\,\E^{\I\omega_0\tau_3}\,,
\quad
m_q = \sqrt{m^{'2} + \mu^2}\,,
\end{equation}
and
\begin{equation}
\chi_0' = 2B_0m'\,, \quad
\chi_3  = 2B_0\mu\,, \quad
\rho    = 2W_0a\,.
\end{equation}
In leading order~(LO) the pion mass is then given by
\begin{equation}
m_\pi^2 = |\chi'| = \sqrt{\chi_0^{\prime2}+\chi_3^2} = 2 B_0 m_q\,.
\end{equation}

A second twist can be introduced in the $(c,s)$-doublet in the same way.
Considering the definition of the fields in Eq.~\eqref{eq:fields}, it should
be noted that the twist in the $(c,s)$-doublet has the opposite sign,
because $s$ and $c$ are exchanged so that they are ordered by increasing
quark masses in the mass matrix.  Following the general argument given
in~\cite{FR1,FR2}, ${\cal O}(a)$-improvement will be realised at twofold
maximum twist, i.\,e.\ when both twist angles equal $\pi/2$. Maintaining
positivity of the fermion determinant requires to implement the twist in the
heavy quark sector orthogonal to the mass splitting between the heavy
quarks~\cite{FR2,Frezzotti}. Because it is more natural to have a diagonal
mass matrix, as e.\,g.\ in~\cite{etmc1}, the twist is implemented
non-diagonally with $\tau_1$.

Referring to the definitions above we define for the light and the heavy
sector
\begin{align}
\chi_{0,l}' &= 2 B_0 m_l' \,, &
\chi_3     &= 2 B_0 \mu_l \,, \\
\chi_{0,h}' &= 2 B_0 m_h' \,, &
\chi_1     &= 2 B_0 \mu_h \,, &
\chi_\delta &= B_0 (m_c-m_s) \,,
\end{align}
where $m_h'$ is the average heavy quark mass. The symmetry breaking
term~$\chi'$ can be written separately for the sectors. With the
corresponding Pauli matrices~$\tau_i$ this reads
\begin{equation}
\chi' =  \left\{
\begin{alignedat}{3}
     &\chi_{0,l}' &\,+\,& \I\chi_3\tau_3  && \text{in}\ (u,d) \,,\\
     &\chi_{0,h}' &\,+\,& \I\chi_1\tau_1 + \chi_\delta\,\tau_3  
                                         &\quad& \text{in}\ (c,s) \,.
\end{alignedat}\right.
\end{equation}
The LO twist angles are $\omega_{0,l}$ and $\omega_{0,h}$, respectively; for
brevity we define $c_{0,l}=\cos(\omega_{0,l})$, $c_{0,h}=\cos(\omega_{0,h})$
and $s_{0,l}=\sin(\omega_{0,l})$, $s_{0,h}=\sin(\omega_{0,h})$ as
in~\cite{ALWW}.

In the following, perturbative quantities denoted by lower case letters
(e.\,g.~$m_\pi^2$) are LO and those with capitals (e.\,g.~$M_\pi^2$) are
NLO.

\section{Meson masses in leading order}

In the framework defined before, we have calculated meson masses and decay
constants in NLO chiral perturbation theory, including lattice artifacts up
to $\mathcal{O}(a^2)$. In this article we shall mainly concentrate on meson
masses, which are of primary interest. Results on the decay constants can be
found in \cite{Sudmann}.

\subsection{Meson masses}

For the meson masses in LO we find the extended Gell-Mann-Okubo mass
formulae
\begin{align}
\label{eq:okubo1}
4m_K^2         &= 3m_{\eta_8}^2 + m_\pi^2\,,\\
\label{eq:okubo2}
m_K^2 + 9m_D^2 &= 6m_{\eta_{15}}^2 + 4m_\pi^2\,,\\
\label{eq:okubo3}
m_K^2 + m_D^2  &= m_{D_s}^2 + m_\pi^2\,,
\end{align}
where the canonical states $\eta_8$ and $\eta_{15}$ are not mass eigenstates
of the theory. For the flavour charged mesons we find in terms of the
underlying parameters
\begin{align}\label{eq:massesLO}
m_\pi^2   & = \sqrt{\chi_{0,l}^{\prime\,2}+\chi_3^2} \,,\nonumber\\
m_{D,K}^2 & = \frac{1}{2}\left( \sqrt{\chi_{0,l}^{\prime\,2}+\chi_3^2}
      + \sqrt{\chi_{0,h}^{\prime\,2}+\chi_1^2} \pm \chi_\delta \right) \,,\\
m_{D_s}^2 & = \sqrt{\chi_{0,h}^{\prime\,2}+\chi_1^2}\,.\nonumber
\end{align}
In Monte Carlo calculations, the renormalised physical PCAC quark masses and
the bare untwisted quark masses can be obtained from the physical and
untwisted lattice currents, respectively, see \cite{etmc1}.  With their
appropriate renormalisation constants on the lattice these full
non-perturbative masses are related to each other by~\cite{etmc1}
\begin{align}
m^\mathrm{PCAC}_l 
 &= Z_P^{-1} \sqrt{(Z_Am^\mathrm{PCAC}_{\chi,l})^2+\mu_l^2} \,,\\
m^\mathrm{PCAC}_{c,s}
 &= Z_P^{-1} \sqrt{(Z_Am^\mathrm{PCAC}_{\chi,h})^2+\mu_\sigma^2} 
          \pm Z_S^{-1} \mu_\delta \,,\nonumber\\
 &= m^\mathrm{PCAC}_h \pm Z_S^{-1} \mu_\delta \,.
\end{align}
In terms of the PCAC quark masses the perturbative Eq.~\eqref{eq:massesLO}
reads
\begin{align}\label{eq:massesLOPCAC}
m_\pi^2   & = 2B_0m^\mathrm{PCAC}_l\,,\nonumber\\
m_{D,K}^2 & = B_0m^\mathrm{PCAC}_l + B_0m^\mathrm{PCAC}_{c,s}\,,\\
m_{D_s}^2 & = 2B_0m^\mathrm{PCAC}_h\,.\nonumber
\end{align}

\subsection{Mixing of flavour neutral mesons}

In general there is mixing between the four flavour-neutral mesons $\pi_3$,
$\eta_8$ and $\eta_{15}$ from the 15-plet and the SU(4)-singlet~$\eta_{1}$,
resulting in the mass eigenstates $\pi^0$, $\eta$, $\eta_c$ and $\eta'$. Due
to isospin symmetry in our case, $\pi_3$ is not mixing with the $\eta$'s. In
a region with ``light'' $s$ and $c$ quarks, where chiral perturbation theory
holds, the $\eta'$ can be integrated out and is not included in ordinary
chiral perturbation theory.

We define the states $\eta$, $\eta_c$ in the mass diagonal basis through
\begin{equation}
  \begin{pmatrix} \eta\\ \eta_c\end{pmatrix}
= \begin{pmatrix} \cos \theta & \sin \theta\\ -\sin \theta &\cos \theta 
     \end{pmatrix} 
  \begin{pmatrix} \eta_8\\ \eta_{15}\end{pmatrix}.
\end{equation}
The mixing angle $\theta$ is in LO given by
\begin{equation}
\tan 2\theta = \sqrt{8}
\left(1+\frac{9}{2}\, \frac{m_D^2-m_K^2}{m_K^2-m_\pi^2} \right)^{-1} \,.
\end{equation}
The masses of the states in the new basis are then given by
\begin{equation}\label{eq:massesLO2}
\begin{aligned}
m_{\eta}^2 &=\, m_{\eta_8}^2 & - \Delta_{s,c}\,, \\
m_{\eta_c}^2 &=\, m_{\eta_{15}}^2 & + \Delta_{s,c}\,,
\end{aligned}
\end{equation}
where $m_{\eta_8}^2$ and $m_{\eta_{15}}^2$ are given by the Gell-Mann-Okubo
mass formulae~\eqref{eq:okubo1}, \eqref{eq:okubo2}, and
\begin{equation}
\Delta_{s,c} = (m_{\eta_{15}}^2-m_{\eta_{8}}^2)\, 
      \frac{\sin^2\theta}{\cos2\theta}
  = \frac{\sqrt2}3 (m_K^2-m_\pi^2) \tan\theta \,,
\end{equation}
which is the analogon to the pion mass splitting in completely
non-degenerate 3-flavour QCD~\cite{gasser}. 

\section{Results in next-to-leading order}

\subsection{Vacuum}

In chiral perturbation theory for twisted mass lattice QCD, the vacuum
configuration is not given by vanishing fields $\phi_a$, corresponding to
$U(x)=\mathds{1}$; instead there is a non-trivial vacuum configuration
$U_0$, which is obtained by minimising the effective action. The physical
fields represent the deviation of $U(x)$ from the vacuum $U_0$. It is
advantageous to parameterise the physical fields through
$U(x)=\sqrt{U_0}\,U_\mathrm{phys}(x)\,\sqrt{U_0}$. In order to calculate
masses and decay constants it is necessary to determine the vacuum $U_0$. In
our case one has
\begin{equation}
U_0 = \exp \left[\frac{\I}{F_0}(\hat\omega_l\lambda_3 + \hat\omega_h\lambda_{13)}\right]\,.
\end{equation}
In NLO we explicitly find
\begin{align}
\hat\omega_l &= \omega_{0,l} 
   - \frac8{F_0^2} \frac{\rho s_{0,l}}{m_\pi^2} 
   \Big\{ m_\pi^2 \, 2W    - m_{D_s}^2\, [4L_6-2W_6]
   + \rho c_{0,l}\,4W' + \rho c_{0,h}\,4[L_6-W_6+W_6'] \Big\} \,,\\
\hat\omega_h &= \omega_{0,h} 
   - \frac8{F_0^2} \frac{\rho s_{0,h}}{m_{D_s}^2} 
   \Big\{ m_{D_s}^2\,2W    - m_\pi^2\,[4L_6-2W_6]
   + \rho c_{0,h}\,4W' + \rho c_{0,l}\,4[L_6-W_6+W_6'] \Big\} \,,
\end{align}
where $W=W_6+W_8/2-2L_6-L_8$ and $W'=W'_6+W'_8/2-W_6-W_8/2+L_6+L_8/2$
(see~\cite{Sharpe-Wu2}).

\subsection{Pion masses}

Calculating the tree-level and one-loop contributions to NLO, we find for
the charged pions
\begin{alignat}{2}
M_{\pi}^2 = m_{\pi}^2 \;&+\;&& \frac{8}{F_0^2} 
  \Big\{ m_{\pi}^4 [2(L_8+2L_6) - (L_5+2L_4)]
      + 2 m_{\pi}^2 m_{D_s}^2 [2L_6 - L_4]\nonumber\\
&-&& \rho\, m_{\pi}^2 c_{0,l} [4(L_8+2L_6) - (L_5+2L_4) - 2(W_8+2W_6) 
     + (W_5+2W_4)]\nonumber\\
&+&& \rho\, m_{\pi}^2 c_{0,h} 2 [L_4-W_4]\nonumber\\
&-&& \rho\,(m_{D_s}^2 c_{0,l} + m_{\pi}^2 c_{0,h}) 2 [2L_6-W_6]\nonumber\\
&+&& \rho^2 c_{0,l}^2 2 [(L_8+2L_6) - (W_8+2W_6) + (W'_8+2W'_6)]\nonumber\\
&+&& \rho^2 c_{0,l} c_{0,h} 4 [L_6 - W_6 + W'_6] \Big\} 
+ \mathrm{loop}_{\pi}.
\end{alignat}
The loop contribution is
\begin{equation}
\mathrm{loop}_{\pi} = \frac{m_{\pi}^2}{N_{\textrm{f}}(4\pi F_0)^2}
\left[ 2 m_{\pi}^2 \ln\left(\frac{m_{\pi}^2}{\Lambda^2}\right)
  - \frac{2+\xi_1}3 m_{\eta}^2 \ln\bigg(\frac{m_{\eta}^2}{\Lambda^2}\bigg)
  - \frac{1-\xi_1}3 m_{\eta_c}^2 \ln\bigg(\frac{m_{\eta_c}^2}{\Lambda^2}\bigg)
\right]
\end{equation}
with $\Lambda = 4\pi F_{\pi}$ conventionally. It depends on
\begin{equation}
\xi_1 \equiv \xi_1(\theta)=\sqrt{2} \sin2\theta-\sin^2\theta\ \in\ [0,1]\,.
\end{equation}
The expression for $M_\pi^2$ explicitly verifies that ${\cal
O}(a)$-improvement of the pion mass requires both twist angles to be at
maximal twist \cite{FR1,FR2}. In particular, a non-maximal twist angle in
the heavy sector would contaminate the pions with ${\cal O}(a)$-terms
proportional to $\rho\, m_\pi^2 c_{0,h}$.

A non-vanishing twist breaks isospin symmetry and leads to a mass splitting
between charged and neutral pions. For the mass splitting we find
\begin{equation}
M_{\pi}^2 - M_{\pi_3}^2
=\frac{16 \rho^2 s_{0,l}^2}{F_0^2}\; [(L_8+2L_6) - (W_8+2W_6) + (W'_8+2W'_6)].
\end{equation}

\subsection{Kaon masses}

The kaons are degenerate at NLO. For their masses we obtain
\begin{alignat}{2}\label{eq:nloKmass}
M_K^2 = m_K^2 \;&+\;&& \frac{4}{F_0^2} \Big\{
     2 m_K^4 [2(L_8+4L_6) - (L_5+4L_4)]
   + 4 m_K^2 (m_D^2 - m_K^2) [2L_6 - L_4]\nonumber\\
&-&& \rho\, m_K^2 (c_{0,l} + c_{0,h}) [4(L_8+4L_6) - (L_5+4L_4)
     - 2(W_8+4W_6) + (W_5+4W_4)]\nonumber\\
&-&& \rho\,(m_D^2 - m_K^2)(c_{0,l} + c_{0,h}) 2 [2L_6-W_6]\nonumber\\
&+&& \rho^2(c_{0,l} + c_{0,h})^2 [(L_8+4L_6) - (W_8+4W_6) + (W'_8+4W'_6)]
  \nonumber\\
&-&& \rho^2(s_{0,l}^2 + s_{0,h}^2) [L_8 - W_8 + W'_8] \Big\} 
+ \mathrm{loop}_K
\end{alignat}
with
\begin{equation}
\mathrm{loop}_K =
\frac{m_K^2}{N_{\textrm{f}}(4\pi F_0)^2}
\left[ \frac{4-\xi_2}{3} m_{\eta}^2 \ln\bigg(\frac{m_{\eta}^2}{\Lambda^2}\bigg)
  -\frac{1-\xi_2}{3} m_{\eta_c}^2 \ln\bigg(\frac{m_{\eta_c}^2}{\Lambda^2}\bigg)
\right]
\end{equation}
and
\begin{equation}
\xi_2 \equiv \xi_2(\theta) 
= 5\sin^2\theta -\sqrt{2} \sin\theta \cos\theta\ 
\in\ \big[\tfrac{5-3\sqrt3}{2},1\big] \,.
\end{equation}
Again, $\mathcal{O}(a)$ improvement is obtained when both light and heavy
twists are maximal. In contrast to the case of pions, however, the kaon
masses contain $\mathcal{O}(a^2)$ terms at maximal twist.

\subsection{Heavy-light case}

The heavy-light case is the situation where the both sectors are mass
degenerate, i.\,e.\ $m_l=m_u=m_d$ and $m_h=m_s=m_c$. In this special case
the mixing angle is analytically given by
\begin{equation}
\tan 2\theta = \sqrt{8}
\end{equation}
This mixing angle corresponds to including the physical states $\eta$ and
$\eta_c$ in the field matrix by replacing the generators $\lambda_8$,
$\lambda_{15}$ by $\lambda_8^\ast$, $\lambda_{15}^\ast$ (see
App.~\ref{generators}).

In LO the mass spectrum given in Eqs.~\eqref{eq:massesLO},
\eqref{eq:massesLOPCAC}, \eqref{eq:massesLO2} simplifies to
\begin{align}
m_{\pi}^2  & =\, 2 B_0 m^{\mathrm{PCAC}}_l \,,\nonumber\\
m_K^2 = m_D^2 = m_{\eta}^2
         & =\, B_0 m^{\mathrm{PCAC}}_h + B_0 m^{\mathrm{PCAC}}_l \,,\\
m_{D_s}^2 = m_{\eta_c}^2
         & =\, 2 B_0 m^{\mathrm{PCAC}}_h \,. \nonumber
\end{align}
The loop contributions simplify too, because of
\begin{equation}
\xi_1(\theta) = \xi_2(\theta) = 1.
\end{equation}

In NLO there are mixing effects between the $K$- and $D$-mesons, when
$m_s=m_c$. In the generic case $m_s \neq m_c$ these mixings contribute only
at NNLO, related to the non-analytic behaviour in $m_s-m_c$. In the case
$m_s=m_c$, however, the mixing already shows up in NLO. In particular it
leads to a flavour breaking in both sectors according to
\begin{equation}
M_{K^\pm}^2 - M_{K^0}^2
=M_{D^\pm}^2 - M_{D^0}^2
=\frac{16\rho^2s_{0,l}s_{0,h}}{F_0^2}\; [ L_8 - W_8 + W'_8 ]\,,
\end{equation}         
and the above formula~(\ref{eq:nloKmass}) for $M_K^2$ gives the average of
the kaon masses. This breaking is a pure lattice artifact.

At tree level the resulting formulae for the kaon masses coincide exactly
with those given in~\cite{ALWW}. In NLO, however, the loop contributions to
the kaon masses
\begin{equation}
\frac{m_K^2}{N_{\textrm{f}}(4\pi F_0)^2}
\bigg[ m_{\eta}^2 \ln\bigg(\frac{m_{\eta}^2}{\Lambda^2}\bigg) \bigg],
\end{equation}
which we obtain, are different from those given in~\cite{ALWW}:
\begin{equation}
\frac{m_K^2}{N_{\textrm{f}}(4\pi F_0)^2}
\bigg[ \frac{4}{3} m_{\eta_8}^2 \ln\bigg(\frac{m_{\eta_8}^2}{\Lambda^2}\bigg)
   - \frac{1}{3} m_{\eta_{15}}^2 
   \ln\bigg(\frac{m_{\eta_{15}}^2}{\Lambda^2}\bigg) \bigg]\,.
\end{equation}
The origin of this discrepancy is the neglect of the mixing of the
$\eta$-mesons in \cite{ALWW}. The results are only valid if the mixing
disappears, which only applies if all quark masses are degenerate.  Also, as
a consequence, the loop-divergences in~\cite{ALWW} do not cancel the
tree-level ones if $m_l\neq m_h$.

A related discrepancy can be found for the kaon decay constant, where our
result for the loop contribution is
\begin{equation}
\frac{N_{\textrm{f}}}{2(4\pi F_0)^2}
\bigg[ \frac{3}{16} m_{\pi}^2 \ln\left(\frac{m_{\pi}^2}{\Lambda^2}\right)
      + \frac{5}{8} m_K^2   \ln\left(\frac{m_K^2}{\Lambda^2}\right)
      + \frac{3}{16} m_{D_s}^2 \ln\bigg(\frac{m_{D_s}^2}{\Lambda^2}\bigg) \bigg]
\end{equation}
in contrast to~\cite{ALWW}:
\begin{equation}
\frac{N_{\textrm{f}}}{2(4\pi F_0)^2}
\bigg[ \frac{3}{16} m_{\pi}^2 \ln\left(\frac{m_{\pi}^2}{\Lambda^2}\right)
      + \frac{1}{2} m_K^2 \ln\left(\frac{m_K^2}{\Lambda^2}\right)
      + \frac{3}{16} m_{\eta_8}^2 \ln\left(\frac{m_{\eta_8}^2}{\Lambda^2}\right)
      + \frac{1}{8} m_{D_s}^2 \ln\bigg(\frac{m_{D_s}^2}{\Lambda^2}\bigg) \bigg] \,.
\end{equation}

\section{Conclusion}

We have obtained the masses of pseudoscalar mesons in the framework of
chiral perturbation theory for twisted mass lattice QCD with a degenerate
doublet of $u$ and $d$ quarks and a non-degenerate doublet of $s$ and $c$
quarks in next-to-leading order, including lattice effects up to
$\mathcal{O}(a^2)$. The results display the dependence of the masses on the
two twist angles for the light and heavy sectors. For maximal twist in both
sectors, Symanzik improvement to $\mathcal{O}(a)$, as proven in
\cite{FR1,FR2}, is explicitly exhibited. The mixing of flavour-neutral
mesons is also discussed. For the case of degenerate $s$ and $c$ quarks,
proper account of mixing corrects results in the literature.

\appendix

\section{Generators of SU($N$)}\label{generators}

The generators of SU($N$) in the fundamental representation are the
hermitian $N \times N$ matrices with zero trace. A useful representation is
as follows \cite{SU4}.  Let $e_{ij}$ be the matrix with the only
non-vanishing entry being 1 in row/column $ij$:
\begin{equation}
(e_{ij})_{kl} = \delta_{ik} \delta_{jl}\,.
\end{equation}
The off-diagonal generators of SU($N$) can be written in terms of $N \times
N$ generalisations of the Pauli matrices $\sigma_1$ and $\sigma_2$, defined
by
\begin{align}
\sigma_{1;ij} &= e_{ij} + e_{ji}\,, &\quad &(i<j), \nonumber\\
\sigma_{2;ij} &= -\I e_{ij} + \I e_{ji}\,, &\quad &(i<j).
\end{align}
The diagonal traceless matrices, generalising $\sigma_3$, are defined by
\begin{equation}
\sigma_{3;i} = \sqrt{\frac{2}{i(i-1)}}\,
\operatorname{diag}(\underbrace{1,\ldots,1}_{i-1},1-i,
   \underbrace{0,\ldots,0}_{N-i})\,, \qquad (i>1).
\end{equation}
The extended Gell-Mann matrices $\lambda_a$ are then given by
\begin{align}
\sigma_{1;ij} &= \lambda_a\,, \quad 
   \text{for} \quad a = (j-1)^2+2i-2\,, \quad (i<j),\nonumber\\
\sigma_{2;ij} &= \lambda_a\,, \quad 
   \text{for} \quad a = (j-1)^2+2i-1\,, \quad (i<j),\\
\sigma_{3;i} &= \lambda_a\,, \quad 
   \text{for} \quad a = i^2-1\,, \quad (i>1),\nonumber
\end{align}
where $a=1,\ldots, N^2-1$, and the generators of SU($N$) are equal to
$\lambda_a /2$. They are orthonormal in the sense of
\begin{equation}
\Tr{\lambda_a \lambda_b} = 2 \delta_{ab}\,.
\end{equation}

For SU(4) the diagonal Gell-Mann matrices, corresponding to the flavour
neutral mesons, are explicitly given by
\begin{align}
\lambda_3         &= \operatorname{diag}(1,-1,0,0) \,, &
\sqrt3\,\lambda_8   &= \operatorname{diag}(1,1,-2,0) \,, &
\sqrt6\,\lambda_{15} &= \operatorname{diag}(1,1,1,-3) \,.
\end{align}

In the heavy-light case the latter two can be replaced for simplicity with
the linear combinations
\begin{align}
\sqrt2\,\lambda_8^\ast &= \operatorname{diag}(1,1,-1,-1) \,, &
\lambda_{15}^\ast    &= \operatorname{diag}(0,0,1,-1) \,.
\end{align}


\end{document}